\documentclass[9pt,twocolumn,twoside]{opticajnl}
\journal{opticajournal} 

\setboolean{shortarticle}{true}

\usepackage{lineno}
\usepackage{upgreek}
\usepackage{stfloats}
\linenumbers 

\title{Cavity-enhanced symmetric second-harmonic generation}

\author[1,2]{Ming-Yuan Gao}
\author[1,2,3,*]{Zhi-Yuan Zhou}
\author[1,2,3,**]{Bao-Sen Shi}

\affil[1]{CAS Key Laboratory of Quantum Information, University of Science and Technology of China, Hefei, Anhui 230026, China}
\affil[2]{CAS Center for Excellence in Quantum Information and Quantum Physics, University of Science and Technology of China, Hefei 230026, China}
\affil[3]{Hefei National Laboratory, University of Science and Technology of China, Hefei 230088, China}

\affil[*]{Corresponding author: zyzhouphy@ustc.edu.cn} 
\affil[**]{Corresponding author: drshi@ustc.edu.cn}

\begin{abstract}
 As one of the two types of backward second-harmonic generation (SHG), symmetric SHG exhibits some physical characteristics and application prospects that are distinct from those of forward SHG. It is generally realized through quasi-phase matching, which imposes more stringent requirements on the poling period and thus presents challenges for domain engineering. Although employing larger poling periods can ease fabrication, it inevitably reduces conversion efficiency, a drawback that can be compensated by using a cavity. In this work, we employed a semi-monolithic cavity to enhance the efficiency of 7th-order symmetric SHG, achieving a measured one-sided conversion efficiency of 7.2\%, which corresponds to a theoretical total efficiency of 14.4\%. This represents an improvement of more than three orders of magnitude compared with the single-pass case. In addition, the nonlinear coefficient of the crystal of ${d_{33}} = 4.8$ $\rm pm/V$ was estimated.
\end{abstract}

\setboolean{displaycopyright}{false} 

\begin{document}
\nolinenumbers
\maketitle

\section{Introduction}
Since second-harmonic generation (SHG) was first demonstrated \cite{1}, related research has driven advancements in nonlinear and quantum optics \cite{2,3,4,5,6}. The basic function of SHG is to double the frequency of the laser, which allows the generation of wavelengths that are difficult or even impossible to produce with lasers. The generated second-harmonic waves have numerous applications, such as pumping an optical parametric oscillator (OPO) to generate classical light fields \cite{5} or photon pairs \cite{6}. Quasi-phase matching (QPM) expands the conditions for second-harmonic generation \cite{7}. Especially through backward QPM (BQPM) \cite{8,9}, two counterpropagating second-harmonic waves can be generated from two counterpropagating \cite{10,11}. As one of the two types of backward SHG, this type of SHG is referred to as symmetric SHG (SSHG), following the term in Ref.\cite{12}. It exhibits several unique characteristics compared with forward SHG (FSHG), where the fundamental and second-harmonic waves co-propagate. In SSHG, the poling period of the nonlinear crystal must satisfy: $\Lambda  = 2\pi m/{k_{2\omega }}$ (which differs from the polarization period requirement for FSHG $\Lambda  = 2\pi m/\left| {2{k_\omega } - {k_{2\omega }}} \right|$ ),  where $m$ is the order of QPM, while ${k_\omega }$ and ${k_{2\omega }}$ are the wave vectors of the fundamental and second-harmonic waves, respectively. The BQPM condition causes it to have a narrower wavelength tuning bandwidth \cite{9}. SSHG involves six different phase-matched processes (in contrast to the two processes in FSHG) \cite{13}. At a finite pump intensity, the conversion efficiency can reach its maximum value (which differs from FSHG where the conversion efficiency asymptotically approaches complete conversion at an infinite pump intensity \cite{10,14}), and below a certain pump intensity, the conversion efficiency is higher than that based on FSHG \cite{10,15}. SSHG also has potential applications in all-optical switching and modulation \cite{15,16}.

Conversion efficiency is a key parameter in SSHG. However, achieving first-order BQPM (i.e., $m = 1$) for SSHG requires a relatively short poling period. Although there have been reports of SSHG and degenerate backward OPOs in the first-order case \cite{12,17}, the sub-micron poling process still faces fabrication challenges. The implementation of higher-order QPM relaxes the requirements for poling periods while introducing the issue of reduced conversion efficiency \cite{18}. A prevalent strategy for efficiency enhancement involves the utilization of a periodically poled crystal waveguide \cite{12,19} or a cavity \cite{18,20,21,22,23,24}. SSHG in waveguides has already been experimentally demonstrated \cite{12}. However, in bulk materials, experimental demonstrations of efficiency enhancement using different types of cavities have primarily focused on FSHG \cite{18,20,21,22,23,24}. To illustrate, blue light at 397.5 nm was generated in a ring cavity by using a periodically poled potassium titanyl phosphate (PPKTP) crystal with a conversion efficiency of 45\% (for the mode-matched fundamental power) \cite{21}. In a ring cavity using a periodically poled lithium niobate crystal, a second-harmonic wave at 776 nm was obtained with two different coupling mirrors yielding conversion efficiencies of 65.8\% and 65.9\% \cite{22}. A second-harmonic wave at 402.5 nm was achieved by using a third-order QPM PPKTP crystal with a conversion efficiency of 15.3\% in a ring cavity \cite{18}. Beyond the ring cavity \cite{18,20,21,22}, monolithic and semi-monolithic standing-wave cavities have also been implemented \cite{23,24}. A second-harmonic wave at 426 nm was achieved with a conversion efficiency of 45\% (with respect to the mode-matched fundamental power) in a monolithic cavity formed by polished and coated both ends of a PPKTP crystal \cite{23}. A second-harmonic wave at 775 nm was generated in a semi-monolithic cavity formed by one polished and coated end of a PPKTP crystal and a cavity mirror, achieving a conversion efficiency of approximately 95\% \cite{24}.  However, to date there have been no experimental reports demonstrating the enhancement of SSHG efficiency through the use of a cavity. Given the inherent requirement for counterpropagating fundamental waves in SSHG, a standing-wave cavity is naturally well-suited for this configuration.

In this work, we demonstrate efficiency enhancement of 7th-order type-0 BQPM SSHG by using a semi-monolithic cavity. With a fundamental power of 50 mW, a second-harmonic one-sided output of 3.6 mW was obtained, corresponding to a one-sided conversion efficiency of 7.2\%. Taking into account the equivalent second-harmonic output from the other side \cite{15}, the total efficiency reaches 14.4\%. This conversion efficiency represents an improvement of more than three orders of magnitude compared with the single-pass SSHG efficiency under the same BQPM condition \cite{9}. Our experiment significantly advances the application potential of SSHG and BQPM in nonlinear and quantum optics.

\section{THEORY MODEL}
\begin{figure}[htb]
\centering\includegraphics[width=8.5cm]{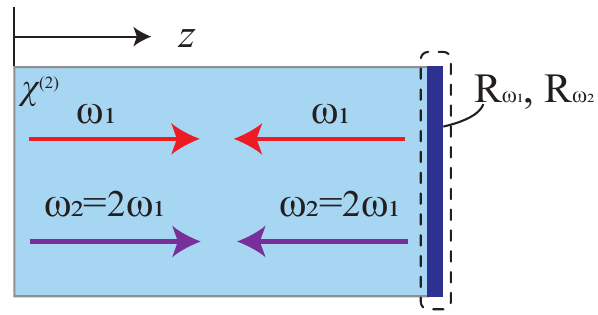}
\caption{Two configurations for SSHG.}
\label{fig1}
\end{figure}

There are two "single-pass" configurations for SSHG, as illustrated in the Fig.\ref{fig1}. The distinction between the two configurations lies in whether the fundamental and/or second-harmonic waves are reflected on one side of the crystal. Due to variations in boundary conditions, the two configurations correspond to distinct conversion efficiency models \cite{15,16}.

Here, we consider the case where the fundamental wave is fully reflected while the second-harmonic wave is not reflected on one side of the crystal (i.e., the reflectivity for the fundamental wave is 1, and for the second-harmonic wave is 0). The corresponding conversion efficiency model is given by \cite{15}:
\begin{equation}
\begin{array}{l}
\eta  = \frac{1}{2} - \frac{1}{2}{\tan ^2}(\frac{\pi }{4} - \sqrt {{\Gamma ^2}{I_1}\eta } )
\end{array}\label{eq1}
\end{equation}
where ${I_1}$ is the intensity of the fundamental wave, $\Gamma $ is given by $\Gamma  = \frac{{4\pi l{d_{eff}}}}{{{\lambda _1}\sqrt {2n_1^2{n_2}{\varepsilon _0}c} }}$, $l$ is the length of the PPKTP crystal, ${d_{eff}}$ is the effective nonlinear coefficient, ${\lambda _1}$ is the wavelength of the fundamental wave, $c$ is the light velocity in vacuum, ${n_1}$ and ${n_2}$ are the refractive indices of the fundamental wave and the second-harmonic wave, respectively, and ${\varepsilon _0}$ is the vacuum dielectric constant. The conversion efficiency in \eqref{eq1} accounts only for the second-harmonic wave propagating in one side. The total efficiency of the second-harmonic waves in both sides is theoretically equal to twice this value \cite{15}. 

The implementation of a cavity significantly enhances the fundamental wave power, thereby enabling substantial improvements in the conversion efficiency. The enhancement factor is given by \cite{25}
\begin{equation}
\begin{array}{l}
\kappa  = \frac{{{P_c}}}{{{P_1}}} = \frac{T}{{{{[1 - \sqrt {(1 - T)(1 - L)(1 - N)} ]}^2}}}
\end{array}\label{eq2}
\end{equation}
where ${P_c}$ is the circulating power established in the cavity from the input power ${P_1}$, $T$ represents the transmittance of the input mirror, $L$ is the round-trip linear loss (without $T$), and $N$ is the nonlinear loss. Here, the nonlinear loss of the cavity corresponds to the conversion efficiency of the second-harmonic wave, which can be expressed as $N = 2\eta $.

The intensity of the fundamental wave ${I_1}$ can be estimated by the expression \cite{26,27}
\begin{equation}
\begin{array}{l}
{I_1} = \frac{{{P_c}}}{{\pi w_1^2}}
\end{array}\label{eq3}
\end{equation}
where ${w_1}$ is the radius of the beam.

By combining \eqref{eq1}, \eqref{eq2}, \eqref{eq3}, and taking into account the effect of mode matching \cite{28}, the conversion efficiency in the cavity is obtained as:
\begin{equation}
\begin{array}{l}
\eta  = \frac{1}{2} - \frac{1}{2}{\tan ^2}(\frac{\pi }{4} - \sqrt {\frac{{{\Gamma ^2}\beta {P_1}\eta T}}{{\pi w_1^2{{[1 - \sqrt {(1 - T)(1 - L)(1 - 2\eta )} ]}^2}}}} )
\end{array}\label{eq4}
\end{equation}
where $\beta $ is the mode-matching coefficient.

\section{EXPERIMENTAL RESULTS}

\begin{figure}[htb]
\centering\includegraphics[width=8.5cm]{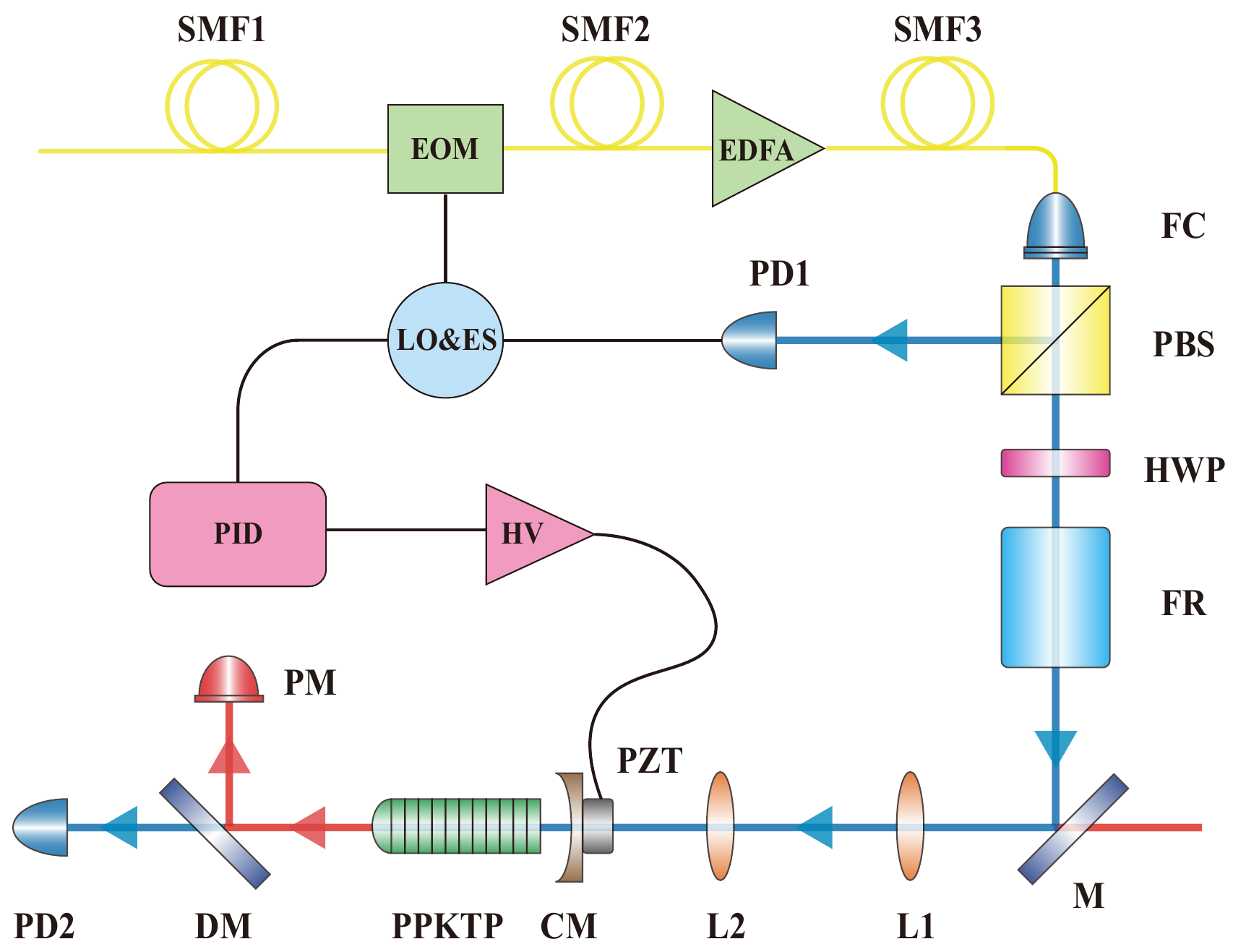}
\caption{Experimental setup. SMF: single-mode fiber; EOM: electro-optic modulator; EDFA: erbium-doped fiber amplifier; FC: fiber coupler; PBS: polarization beam splitter; HWP: half-wave plate; FR: Faraday Rotator; M: mirror; L: lens; PZT: piezoelectric transducer; CM: cavity mirror; PPKTP: periodically poled KTP crystal; DM: dichroic mirror;  PM: power meter; PD: photodetector; LO\&ES: local oscillator and error signal generator; PID: proportional–integral–derivative controller; HV: high-voltage amplifier.}
\label{fig2}
\end{figure}

The experimental setup is shown in the Fig.\ref{fig2}. A continuous-wave (CW) laser at 1556.8 nm is coupled into a single-mode fiber and undergone phase modulation by an EOM to generate sidebands, which are used to produce the Pound–Drever–Hall \cite{29} error signal for cavity length stabilization. The modulation frequency is controlled by the LO. 
The laser is then amplified by an EDFA and output through an FC to pump the SSHG cavity. The polarization of the output laser is adjusted to vertical (V) after passing through the PBS, HWP, and FR. Additionally, the HWP and FR adjust the polarization of the light reflected back from the cavity to V, allowing it to be reflected by the PBS and detected by PD1 to obtain the reflection spectrum of the cavity. This spectral signal also serves as an input for generating the error signal. The error signal is sent to the PID circuit, which generates a control signal that is then amplified by the HV. This amplified signal drives a PZT attached to the cavity mirror to stabilize the cavity length. Two lenses, L1 and L2, with focal lengths of 75 mm and 50 mm, respectively, are used for mode matching. 
The input mirror of the cavity has a nominal reflectivity of approximately 99.2\% for the fundamental wave and a concave curvature radius of 15 mm.
The other cavity mirror is formed by the convex surface of a PPKTP crystal, which has a curvature radius of 8 mm and a reflectivity exceeding 99.9\% for the fundamental wave.
The PPKTP crystal has dimensions of $1 \times 2 \times 9$ $\rm m{m^3}$ and a poling period of 2.95 µm, satisfying the 7th-order type-0 BQPM condition. The temperature of the crystal was maintained at 25°C. Both cavity mirrors have high transmittance for the second-harmonic wave, allowing second-harmonic output from both sides of the cavity.
The leftward-propagating second-harmonic (LSH) wave and the leaked fundamental wave from the cavity mirror are separated by the DM. The LSH power is measured using the PM, while the detector PD2 records the transmission spectrum of the cavity.

\begin{figure*}[ht]
\centering\includegraphics[width=13cm]{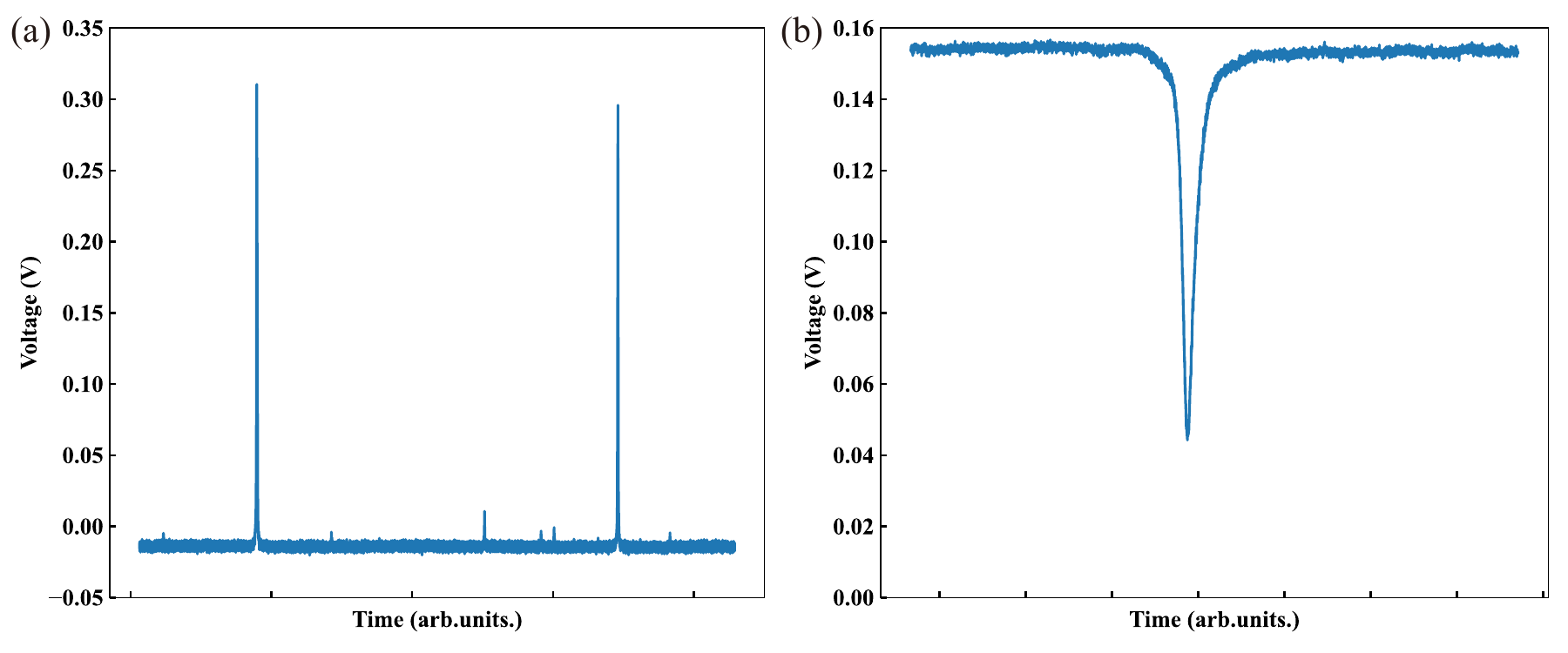}
\caption{(a) Reflection and (b) transmission spectrum of the cavity.}
\label{fig3}
\end{figure*}

By scanning the cavity length, the cavity's reflection and transmission spectra were recorded using an oscilloscope, as shown in the Fig.\ref{fig3}. Here, the fundamental wave power was approximately 10 mW. From the transmission spectrum, the finesse was determined to be approximately 619, and using the formula $\gamma  \approx 2\pi /F$, the total loss $\gamma $ was estimated to be approximately 1\%. The estimated transmittance of the input mirror was 0.7\%, resulting in the remaining loss of 0.3\%. In this case, the impedance-matching coefficient was determined to be approximately 84\% in theory \cite{25}. Once the cavity is locked, the transmittance of the input mirror is determined by the product of the impedance-matching coefficient and the mode-matching coefficient, which can be calculated from the reflection spectrum of the cavity \cite{28}. Therefore, the mode-matching coefficient of the cavity $\beta  \approx 85\% $ can then be obtained.

\begin{figure}[htb]
\centering\includegraphics[width=8.5cm]{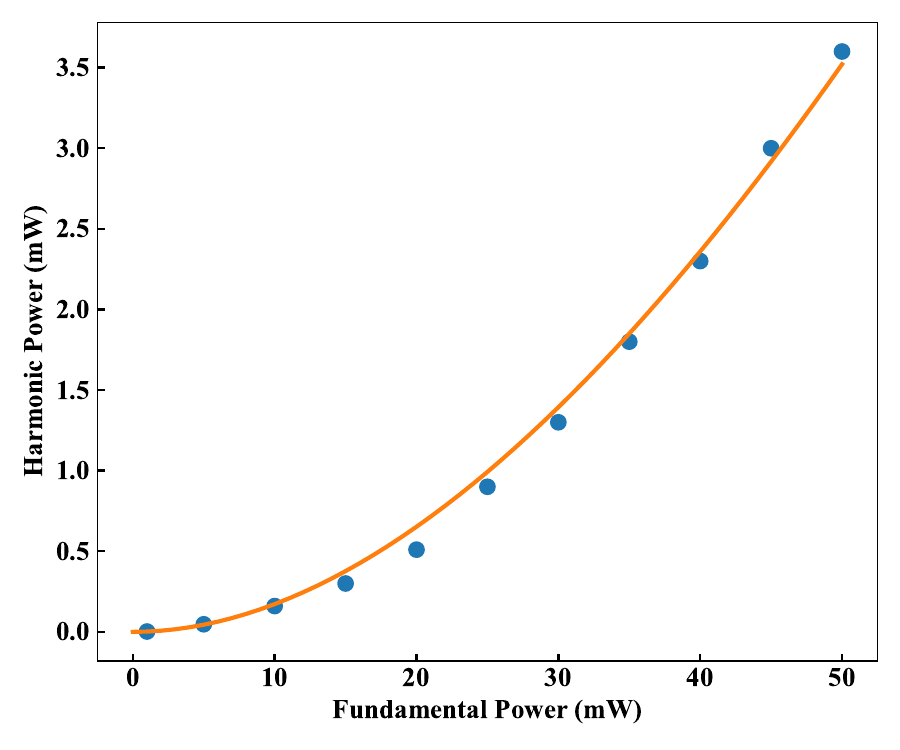}
\caption{Leftward-propagating second-harmonic power as a function of the fundamental power. The dots represent the experimental data and the curve shows the results of the simulation.}
\label{fig4}
\end{figure}

The temperature of the crystal was maintained at 25°C. By tuning the laser wavelength and locking the cavity length, the wavelength that maximized the LSH conversion efficiency was determined to be 1556.8 nm. The LSH power was measured for varying the power of the fundamental wave, as shown in the Fig.\ref{fig4}. With the fundamental power of 50 mW, the second-harmonic power was measured to be 3.6 mW, corresponding to a conversion efficiency of 7.2\%. The one-sided conversion efficiency exhibits an enhancement of more than $4 \times {10^3}$ compared with the one-sided efficiency of single-pass SSHG under the same order and type of BQPM (but with the different length of the crystal, where the length of 4.5 mm was used in the single-pass configuration) \cite{9}. Considering the equal power of LSH and rightward-propagating second-harmonic generation, the total conversion efficiency is 14.4\% in theory.

Next, using \eqref{eq4}, we provide an estimated value for the nonlinear coefficient ${d_{33}}$. The refractive index is obtained from the Ref.\cite{30}, the beam radius is estimated to be 35 µm, and the effective nonlinear coefficient in the current experiment is ${d_{eff}} = 2{d_{33}}/7\pi $ \cite{31}. The above values and equation are substituted into \eqref{eq4}, and when ${d_{33}} = 4.8$ $\rm pm/V$, the solid line in the Fig.\ref{fig4} is obtained. Previously, Ref.\cite{32} reported first-order QPM FSHG in PPKTP with the same poling period as in this work, where ${d_{33}} = 8.6~\mathrm{pm/V}$. It is worth mentioning that higher-order QPM has a smaller tolerance for periodic fabrication errors, which can lead to a decrease in efficiency \cite{31,33}. As a result, under otherwise identical conditions, the estimated value of ${d_{33}}$ may be lower than that in the case of first-order QPM.

\section{Conclusion}
In conclusion, we experimentally demonstrate the generation of SSHG in a semi-monolithic cavity. Utilizing a 7th-order type-0 BQPM PPKTP crystal, we achieve a one-sided conversion efficiency of 7.2\% under the CW pumping, corresponding to a total conversion efficiency of 14.4\% in theory. This efficiency marks a significant improvement over the previously reported single-pass case. This work further paves the way for advancing SSHG and BQPM applications in nonlinear and quantum optics.

\begin{backmatter}
\bmsection{Funding}
 National Key Research and Development Program of China (2022YFB3903102, 2022YFB3607700), National Natural Science Foundation of China (NSFC)(62435018), Innovation Program for Quantum Science and Technology (2021ZD0301100), USTC Research Funds of the Double First-Class Initiative(YD2030002023), and Research Cooperation Fund of SAST, CASC (SAST2022-075).

\bmsection{Disclosures}
The authors declare no conflicts of interest.

\bmsection{Data availability}
Data underlying the results presented in this paper are not publicly available at this time but may be obtained from the authors upon reasonable request.

\end{backmatter}

\bibliography{References}

\end{document}